# Impaired neurovascular coupling in the APPxPS1 mouse model of Alzheimer's disease

Armelle Rancillac*, Hélène Geoffroy and Jean Rossier

Laboratoire de Neurobiologie, CNRS UMR 7637, ESPCI ParisTech, Paris, France.

**Running title**: Impaired neurovascular coupling in AD.

* Correspondence to be addressed to:

Dr. Armelle Rancillac,

Laboratoire de Neurobiologie,

CNRS UMR 7637, ESPCI ParisTech,

10 rue Vauquelin, 75005 Paris, France.

Tel.: +33 (0)1 40 79 51 83

Fax: +33 (0)1 40 79 47 57

E-mail: armelle.rancillac@college-de-france.fr

**Abstract**

The tight coupling between neuronal activity and the local increase of blood flow termed neurovascular coupling is essential for normal brain function. This mechanism of regulation is compromised in Alzheimer's Disease (AD). In order to determine whether a purely vascular dysfunction or a neuronal alteration of blood vessels diameter control could be responsible for the impaired neurovascular coupling observed in AD, blood vessels reactivity in response to different pharmacological stimulations was examined in double transgenic APPxPS1 mice model of AD.

Blood vessels movements were monitored using infrared videomicroscopy *ex vivo*, in cortical slices of 8 month-old APPxPS1 and wild type (WT) mice. We quantified vasomotor responses induced either by direct blood vessel stimulation with a thromboxane $A_2$ analogue, the U46619 (9,11-dideoxy-11a,9a-epoxymethanoprostaglandin F2α) or via the stimulation of interneurons with the nicotinic acetylcholine receptor (nAChRs) agonist DMPP (1,1-Dimethyl-4-phenylpiperazinium iodide). Using both types of stimulation, no significant differences were detected for the amplitude of blood vessel diameter changes between the transgenic APPxPS1 mice model of AD and WT mice, although the kinetics of recovery were slower in APPxPS1 mice. We find that activation of neocortical interneurons with DMPP induced both vasodilation via Nitric Oxide (NO) release and constriction via Neuropeptide Y (NPY) release. However, we observed a smaller proportion of reactive blood vessels following a neuronal activation in transgenic mice compared with WT mice. Altogether, these results suggest that in this mouse model of AD, deficiency in the cortical neurovascular coupling essentially results from a neuronal and not a vascular dysfunction.

**Introduction**

Alzheimer's Disease (AD) is a multifactorial and progressive neurodegenerative pathology associated with various vascular dysfunctions. During AD development, levels of beta amyloid (Aβ) peptides increase and form depositions in neuritic plaques and in cerebral blood vessels, causing cerebral amyloid angiopathy (CAA). Disturbances in neuronal circuits, such as loss of neurons and synapses [1-3] as well as progressive vascular dysregulation [4] are well known features of this pathology.

Neurons and astrocytes are known to function in a close metabolic and structural interdependence with blood vessels forming the so called neurovascular unit [5,6]. The tight coupling between increased neuronal activity and local control of cerebral blood vessels diameter known as neurovascular coupling or functional hyperemia is precociously altered in AD patients [7-10] and in various AD transgenic mouse models [4,11-13]. In AD, ultrastructural changes affecting diverse properties of the neurovascular unit have been observed, such as capillary distortions, blood-brain-barrier (BBB) dysfunction, losses of endothelial mitochondria, thickening of the basement membrane, endothelial degeneration, increase in degererative pericytes, perivascular neuroinflammation [14]. Moreover, early in the physiopathological development of the disease, the AD brain is characterized by an altered perfusion to activated areas [15]. These ultrastructural changes affecting the neurovascular unit result in disturbed regulatory mechanisms and contribute to cerebral damage.

This neurovascular dysfunction observed in AD could result from purely vascular alterations or from a neurodegenerative process [16-18] and a progressive loss of the neuronal control of blood vessels tonus.

Using pharmacological stimulation of blood vessels within cortical slices, the objective of this study was to determine *ex vivo*, the vascular or neuronal origin of the impaired neurovascular coupling occurring during AD. To this aim, here we used 8 month-old double transgenic mice overexpressing human mutated genes of both human amyloid precursor protein (APP751 with ~~both~~ the Swedish and London mutations; SL) and humanpresenilin 1 (PS1M146L) [19]. This APPxPS1 mouse model of AD develops early-onset brain amyloidosis (3-4 months), undergoes substantial cell loss and CAA could first be detected from about 20 weeks of age. We performed direct stimulation of blood vessels using a thromboxane $A_2$ analogue, the U46619 (9,11-dideoxy-11a,9a-epoxymethanoprostaglandin F2α) acting on G-protein coupled thromboxane receptor located on endothelial cells of blood vessels [20], or indirect stimulation via neuronal activation using the nicotinic acetylcholine receptor (nAchRs) agonist DMPP (1,1-Dimethyl-4-phenylpiperazinium iodide). Indeed, in the neocortex, the neurovascular coupling is in part mediated by a direct interaction with interneurons [5,21,22] via the release of vasoactive messengers. As functional postsynaptic nicotinic acetylcholine receptors (nAchRs) are only expressed by non Fast Spiking (FS) neocortical inhibitory interneurons [23-25], DMPP allows a specific neuronal activation.

Here, we observed that for both direct and indirect stimulation protocols, the amplitudes of evoked blood vessel movements were not significantly different in APPxPS1 versus WT mouse. We find that activation of neocortical interneurons with DMPP could dilate blood vessels via Nitric Oxide (NO) or constrict blood vessels via Neuropepide Y (NPY) release. However, the neuronal stimulation using DMPP induced less vasomotor responses in APPxPS1 compared to WT mice. These findings suggest that a loss of neuronal control rather than a purely vascular impairment underlie the neurovascular alteration occurring during the development of AD.

**Material and Methods**

*Animals*

We used nine wild type $C_{57}Bl_6J$ (WT) and six APPxPS1 transgenic 8 month-old male mice, as previously described [19]. Transgenic mice (Tg) were generated by crossing homozygous PS1 (carrying presenilin 1 M146L mutation under the control of the HMG-CoA reductase promoter) with hemizygous APPSL mice (hAPP751 transgene with the Swedish K670N/M671L and London V717I mutations, under the control of the Thy1 promoter) to obtain APPxPS1 transgenics.

All animals were housed in a temperature-controlled (21–25°C) room under daylight conditions. They arrived in the laboratory 2 weeks before initiating experiments to acclimate to their new environment. Experimental procedures were conducted in strict compliance with approved institutional protocols and in accordance with the provisions for animal care and use described in the European Communities Council directive of 24 November 1986 (86-16-09/EEC). Experiments were performed in compliance and following approval of the Sanofi-Aventis Animal Care and use Committee.

*Slice preparation*

Mice were anesthetized using halothane and decapitated. Brains were quickly removed and placed into cold (~4 °C) cutting solution containing (in mM): 110 choline chloride, 11.6 Na-ascorbate, 7 $MgCl_2$, 2.5 KCl, 1.25 $NaH_2PO_4$, and 0.5 $CaCl_2$, continuously bubbled with 95% $O_2$ – 5% $CO_2$. Coronal brain slices (300 µm thick) comprising the somatosensory cortex were cut with a vibratome (VT1200S; Leica, Nussloch, Germany), and transferred to a holding chamber filled with artificial cerebrospinal fluid (ACSF), containing (in mM): 126 NaCl, 2.5 KCl, 1.25 $NaH_2PO_4$, 2 $CaCl_2$, 1 $MgCl_2$, 26 $NaHCO_3$, 20 glucose, and 1 kynurenic acid (a nonspecific glutamate receptor antagonist, Sigma, France), constantly oxygenated (95% $O_2$ – 5% $CO_2$) and held at room temperature. Individual slices were next placed in a submerged recording chamber kept at 32°C and perfused (1.5 ml/min) with oxygenated ACSF (in the absence of kynurenic acid). Blood vessels were visualized in the slice using infrared videomicroscopy with Dodt gradient contrast optics.

*Drugs*

Thromboxane $A_2$ receptors and nicotinic acetylcholine receptors agonists (9,11-dideoxy-11a,9a-epoxymethanoprostaglandin F2α, U46619, 50 nM and 1,1-Dimethyl-4-phenylpiperazinium iodide, DMPP, 100 µM, respectively, Sigma, France) were added in the ACSF perfusion of the cortical slices for 6 min. To block the NOS, slices were treated for at least 1 h with an irreversible inhibitor of constitutive nitric oxide synthase (nNOS) and a reversible inhibitor of inducible nitric oxide synthase (iNOS), Nω-Nitro-L-arginine (L-NNA; 100 µM; Sigma, France). NPY Y1 receptors were blocked with the selective antagonist N2-(Diphenylacetyl)-N-[(4-hydroxyphenyl)methyl]-D-arginine amide (BIBP 3226; 100 µM; Sigma, France).

*Vascular reactivity*

Blood vessel movements were monitored at 32°C by infrared videomicroscopy on cortical slices. Blood vessels remaining in focal plane for more than 50 µm, exhibiting a well-defined luminal diameter (10-20 µm) and located in supragranular layers of the somatosensory cortex were selected for vascular reactivity. Images were acquired every 15 s using Image Pro Plus 7.0 (MediaCybernetics, San Diego, CA) and control baselines of blood vessel diameters were determined for 5 min. Blood vessels with unstable baseline were discarded from the analysis. U46619 or DMPP were applied during 6 minutes after stable baseline period. Luminal diameters were quantified off-line at different locations along the blood vessel using custom written routines running within IgorPro (Wavemetrics) to determine the most reactive part of the vessel. Maximal vasomotor responses were then expressed as percentages of the mean baseline diameter during the control period.

*Statistical analyses*

To determine the statistical significance of the vasomotor responses, changes in diameter were compared using a paired *t*-test at the time before the onset of the drug application where the mean diameter change is closest to 0% and at the time where the maximal response was observed. Vascular reactivities in control WT versus APPxPS1 mice were compared using *t*-test. To compare the proportions of reactive blood vessels in WT versus APPxPS1, we used a z test. The Area Under the Curve (AUC) for the vascular response of each blood vessel was calculated using the trapezoidal rule. For U46619 study, the AUC was calculated as the incremental or decremental values from the baseline obtained during the 27 minutes (from time 3 to 30 minutes after U46619 onset application). For DMPP study, the AUC was calculated as the incremental or decremental values from the baseline obtained during the 12 minutes (from time 0 to 12 minutes after DMPP onset application). Comparisons of AUC among genotype groups were performed by Mann Whitney U test. The AUC was expressed in arbitrary units.

*Histology and Immumohistochemistry*

Following the recordings of blood vessel reactivity, slices were fixed over night. Amyloid deposits were labeled by standard Congo red staining [26], and amyloid loads were manually delineated and quantified with Photoshop CS3. A total of 2-3 measures were performed and averaged for every mouse. Evaluation of amyloid loads was performed in the barrel somatosensory cortex and in the CA1 region of the hippocampus. Blood vessels in transgenic versus WT mice were assessed by immunostaining for basement membrane protein collagen-IV (goat anti-collagen type IV; 1: 400; Chemicon). Blood vessels density was measured using computer-based thresholding methods in Photoshop CS3. A total of 2-3 measures were performed and averaged for every mouse in the barrel somatosensory cortex.

## Results

**Amyloid deposition and vascular density in 8 month-old APPxPS1 mice**

In order to examine the vascular reactivity in the somatosensory cortex of 8 months APPxPS1 versus WT mice, we first looked at the amyloid depositions and the vascular density in this region of interest. Neuropathological analyses revealed in APPxPS1 the presence of numerous extracellular aggregated (Congo red positive) Aβ deposits (Fig. 1A-C) whereas none could be observed in WT mice. Interestingly, we observed that Aβ deposits in the cortex were more numerous and bigger in deep layers, compared to supragranular layers (Fig. 1A and C). Some Congo red positive blood vessels were observed (Fig. 1B), indicating a cerebral amyloid angiopathy at 8 months in this APPxPS1 mice. We quantified Congo red positive aggregates in the cortex and the CA1 field of the hippocampus. Amyloid loads represented 4.53 ± 0.57% of the parenchymal tissue in the somatosensory cortex, and 5.41 ± 0.66% in the CA1 region (no significant difference between these regions) (Fig. 1E).

In the cortex, immunostaining of collagen-IV to label blood vessels revealed no difference in the mean vascular density in APPxPS1 (12.67 ± 1.48%) compared to the WT mice (15.15 ± 2.61%) (Fig. 1 D and F).

**Direct stimulation of blood vessels using U46619**

To determine if purely vascular dysfunctions or a loss of neural control underlay the neurovascular coupling impairment observed in AD, we examined blood vessels reactivity in 8 months APPxPS1 versus WT mice. Using infrared videomicroscopy on cortical slice preparations, we first investigated a potential vascular dysfunction by using direct pharmacological stimulation of blood vessels. We applied a thromboxane A2 stable analogue, the U46619 (50 nM) for 6 minutes on acute WT or APPxPS1 cortical slices. We focused our study on penetrating arterioles that are known to be of prime importance in feeding deeply located microvessels and neurons [27]. Well defined arterioles of supragranular layers were selected for quantitative analyses. This direct vascular stimulation induced strong, significant and reversible vasoconstrictions in both WT (80.82 ± 6.59%, n = 8; $p<0.05$) and APPxPS1 mice (80.83 ± 3.79%, n = 10; $p<0.001$) (Fig. 2, Table 1). The diameter change measured either at 14 minutes after the onset of U46619 application, where the maximum of the response was reached, or at 30 minutes after the onset of U46619 application, did not differ significantly between APPxPS1 and WT mice. However, when the response of each arteriole to the U46619 application was expressed as the area under the response curve (AUC), the percent AUC for the U46619–induced vasoconstriction was significantly highest at 30 minutes in APPxPS1 mice compared to WT mice (325.15 ± 47.31 and 183.94 ± 37.91 respectively, P<0.05, Table 1). Finally, the proportion of blood vessels that constrict in response to U46619 in APPxPS1 mice (10 out of 18) compared to WT mice (8 out of 14) did not differ significantly (p>0.68, Table 1). These results suggest that, at 8 months of age, blood vessels recovery following a direct stimulation is slower but reactivity remains functional in these transgenic mice.

**Indirect stimulation of blood vessels via neuronal activation using DMPP**

Next, in order to investigate whether neuronal release of vasoactive compounds was altered in APPxPS1 mice, we used the nicotinic receptor agonist DMPP (100 µM) to stimulate interneurons. DMPP induced, both in WT or APPxPS1 mice, either vasodilatations (105.21 ± 1.21%, n = 12 and 106.08 ± 2.91%, n = 3, respectively), or vasoconstrictions (95.52 ± 1.08%, n = 7 and 93.42 ± 4.77%, n = 4, respectively; Table 1, Fig. 3). Here again, no significant differences (p>0.56) were observed for the mean vasodilatation or vasoconstriction amplitudes between both lines of mice. However, the proportion of blood vessels reactive to DMPP was smaller in APPxPS1 (7 out of 20) compared to WT (19 out of 27) mice (p<0.05, z-test). In particular, vasodilation were less frequently induced by DMPP in transgenic mice (3 out of 20; 15%) compared to WT mice (12 out of 27; 44%; Table 1). Following DMPP stimulation, proportion of vasoconstriction in transgenic mice (7 out of 27; 26%) was not significantly different compared to WT mice (4 out of 19, 20%; z-test, P>0.8; Table 1). However, no significant difference of AUC for DMPP-induced vasodilation or vasoconstriction were observed in APPxPS1 mice compared with WT mice (P>0.64, Table 1).

These results suggest that the impairment of the functional hyperemia observed in AD is primarily due to a defect in the neuronal control of blood vessels rather than to a pure vascular dysfunction.

**Vasodilations induced by DMPP application are mediated by NO whereas vasoconstrictions are due to NPY release**

The stimulation of GABAergic interneurons can induce vasomotricity via the production of vasoactive substances [5,21,22]. DMPP depolarizes nicotinic cholinergic receptors (nAChR)-expressing interneurons that were shown to be also responsive to serotonergic via the serotonin 5-hydroxytryptamine 3A (5-HT$_{3A}$) receptor [28]. As the pharmacological stimulation of this receptor was recently reported to bi-directionally control arterioles diameter by the release of NO to dilate, or NPY to constrict [29], we hypothesized that DMPP induced vasomotor changes could also be due to NO or NPY release.

Therefore, in order to determine the molecular events underlying vasomotor changes following DMPP stimulation, we successively blocked different possible mechanisms. Lowering basal NO levels by treatment with the constitutive nNOS inhibitor L-NNA (100 µM) prevent all vasodilations in response to DMPP applications. Only vasoconstrictions were recorded (94.65 ± 2.41 of baseline diameter; n = 7/16, P<0.01) (figure 4A). These results suggest that DMPP-induced vasodilations are mediated by NO release.

Then, to determine the molecular pathway underlying vasoconstrictions, treatment of DMPP was reproduced in the presence of NPY Y1 receptor antagonist (BIBP 3226, 1 µM). Indeed, vasoconstrictions mediated by NPY are known to be mediated by smooth muscle NPY Y1 vascular receptor [30]. Under BIBP 3226, constrictions where blocked and only dilations (106.24 ± 2.02%; n = 5/13; P<0.01) could be recorded (figure 4B). Amplitudes of vasomotor responses under these different conditions were not statistically different from control condition. Altogether, these data strongly suggest that pharmacological stimulations of nAchRs-expressing interneurons induce vasodilations through NO release, whereas they induce vasoconstriction through NPY release and activation of its Y1 receptor.

**Discussion**

The alteration of the neurovascular coupling occurring during AD, could result either from a purely vascular disorder or from a neurodegenerative process leading to a loss of blood vessels control by neuronal activity. Indeed, increasing evidence suggests a causal role for vascular disorder in the development of AD. Vascular risk factors including hypertension, diabetes, and hypercholesterolemia increase the risk of incident AD dementia [31] and regional cerebral hypoperfusion is one of the earliest pathological features of AD. On the other hand, various modes of neurodegeneration can coexist in a same mouse model and lead to reduced blood vessels innervation [32]. Previous findings already shown that Aβ accumulates in neurons [33-36], as it was also observed at 2 months in this model [37], which might cause a synaptic degeneration [1] and lead to neuronal death [38]. Extracellular amyolid-β induces significant entorhinal neuronal loss of principal and SOM/NPY neurons in these APPxPS1 mice [39]. Moreover, autophagy is induced but impaired in affected neurons in the AD brain [40,41], causing autophagic vacuoles to accumulate profusely in dystrophic neuritis. This axonopathy disturbs neuronal trafficking [42] and could perturb vasomotor peptidergic release.

**Vascular dysfunction**

In the present study, to determine if purely vascular or neuronal dysfunction were responsible for the decreased neurovascular coupling occurring during AD, we used either direct blood vessels stimulation, or neuronal activation, while recording vasomotor changes. Here, no differences could be observed in the peak amplitude of evoked blood vessels movements between WT and APPxPS1 mice using either type of stimulation. However, the AUC for the U46619-induced vasoconstriction was significantly highest in APPxPS1 mice compared to WT mice, accordingly to *in vivo* studies that reported in other models of AD that vasoconstrictions are usually exaggerated [43-45]. This suggests reduced vessel elasticity, which could be due to CAA. Indeed, CAA was first detected from about 20 weeks of age and developed beyond the age of 40 weeks in this mouse line, with a high prevalenceof the $A\beta_{40}$ isoform forming amyloid deposits in the vascular system [46]. The blood a vessel inelasticity could compromise mental function by preventing neural regions from reacting to new activity by adapting blood supply [47].

**Neuronal dysfunction**

Here, we observed that following a neuronal stimulation, reactive vessels were less numerous in transgenic mice compared to WT mice. Therefore, it is likely that neurodegeneration rather than a purely vascular impairment could be responsible for the alteration of neurovascular coupling in AD.

Indeed, a neuronal loss in the entorhinal cortex [39] and an extensive, selective and early neurodegeneration of the dendritic inhibitory interneurons of the hippocampus have been observed in this model. At 6 months of age, a diminution of 50-60% of SOM-immunopositive neurons was noted [48], in accordance with the loss of SOM and/or NPY neurons frequently reported in AD patients and the linear correlation between SOM and/or NPY deficiency and $A\beta$ content which have been reported [48]. As extracellular $A\beta$ deposits are observed in the hippocampus and in the neocortex (Fig. 1) and [37], this suggest that a neurodegeneration of cortical interneurons could be responsible for the decreased proportion of reactive blood vessels following a neuronal stimulation observed in this study.

Alternatively, Alzheimer's disease is characterized by large losses of nicotinic cholinergic receptors [49,50], that could also contribute to the lack of reactivity following DMPP application.

**Dual role of nAChR-expressing interneurons**

Our results indicate that activation of nAChR-expressing interneurons induces a complex vascular response. Indeed the selective nAChR agonist DMPP induced either constrictions (37%) or dilations (63%) of penetrating arterioles within supragranular layers. All vasoconstrictions were abolished in the presence of the NPY receptor antagonist (BIBP 3226), suggesting that they were elicited by NPY release. This result confirm and extend prior studies implicating NPY in vasoconstrictions [21,51,52].

Conversely, DMPP-induced dilations were blocked in the presence of a nNOS inhibitor suggesting that NO is the predominant messenger inducing the vasodilation, in line previous studies performed in the cerebellum [22,53,54] and in the cortex [21,55-59]. This result confirm the previous observation that nAChRs activation increased levels of the metabolites of NO, and therefore presumably of NO [60]. Together, our observations that in the somatosensory cortex a selective nAChR agonist induce NO mediated vasodilations confirms and extends prior studies reinforcing the central role of NO in the neurovascular coupling.

**Role of astrocytes in the vascular response to DMPP**

Astrocytes are known as cellular intermediaries that couple neuronal activity to local blood flow changes through the phospholipase A2 (PLA2)-mediated synthesis of arachidonic acid, which leads to production of prostaglandins and epoxyeicosatrienoic acids [61].

In the present study, we demonstrated that following DMPP stimulation, vasoconstrictions were mediated by NPY release. This neuropeptide is released by interneurons and directly activates NPY Y1 receptors located on vascular smooth muscle cells. Otherwise, we demonstrated that following DMPPP stimulation, vasodilations were mediated by NO release. NO is known to either directly acts on vascular smooth muscle cells through cGMP-PKG [62,63] or indirectly interacts with numerous signaling pathways also involved in the vascular control [64-66]. In particular, NO could induce astrocytic cytosolic $Ca^{2+}$ increased [67]. However, *in vivo* data strongly suggest that the involvement of astrocytes is likely to occur in the late phase of the neurovascular coupling [68,69]. Even if further experiments would be necessary to fully characterize the NO mediated vasodilations following DMPP application, it appears that in our stimulation conditions, no other components possibly released by astrocytes are involved in the neurovascular coupling.

**Conclusion**

In this study, we observed by infrared videomicroscopy, the cortical vascular response induced by either a vascular activator, the thomboxane A2 analogue, or by a neuronal activator, the nicotinic agonist DMPP in an APPxPS1 transgenic mouse model of AD. No differences were observed in the amplitude of vascular responses between WT and transgenic mice following both types of activators. Therefore, in this mouse model of AD a neuronal alteration rather than vascular changes seems responsible for the impaired neurovascular coupling.

**Acknowledgments**

We thank the Sanofi-Aventis Neurodegenerative Disease Group for the generous gift of the animals involved in this study and Marcel Leopoldie for animal husbandry. We are grateful to Dr. Isabelle Férézou for helpful comments reading the manuscript and to Marion Daenens and Thierry Dendele for technical support. This work was supported by the French National Research Agency (ANR-06-NEURO-033-01grant).

**Figure Legends**

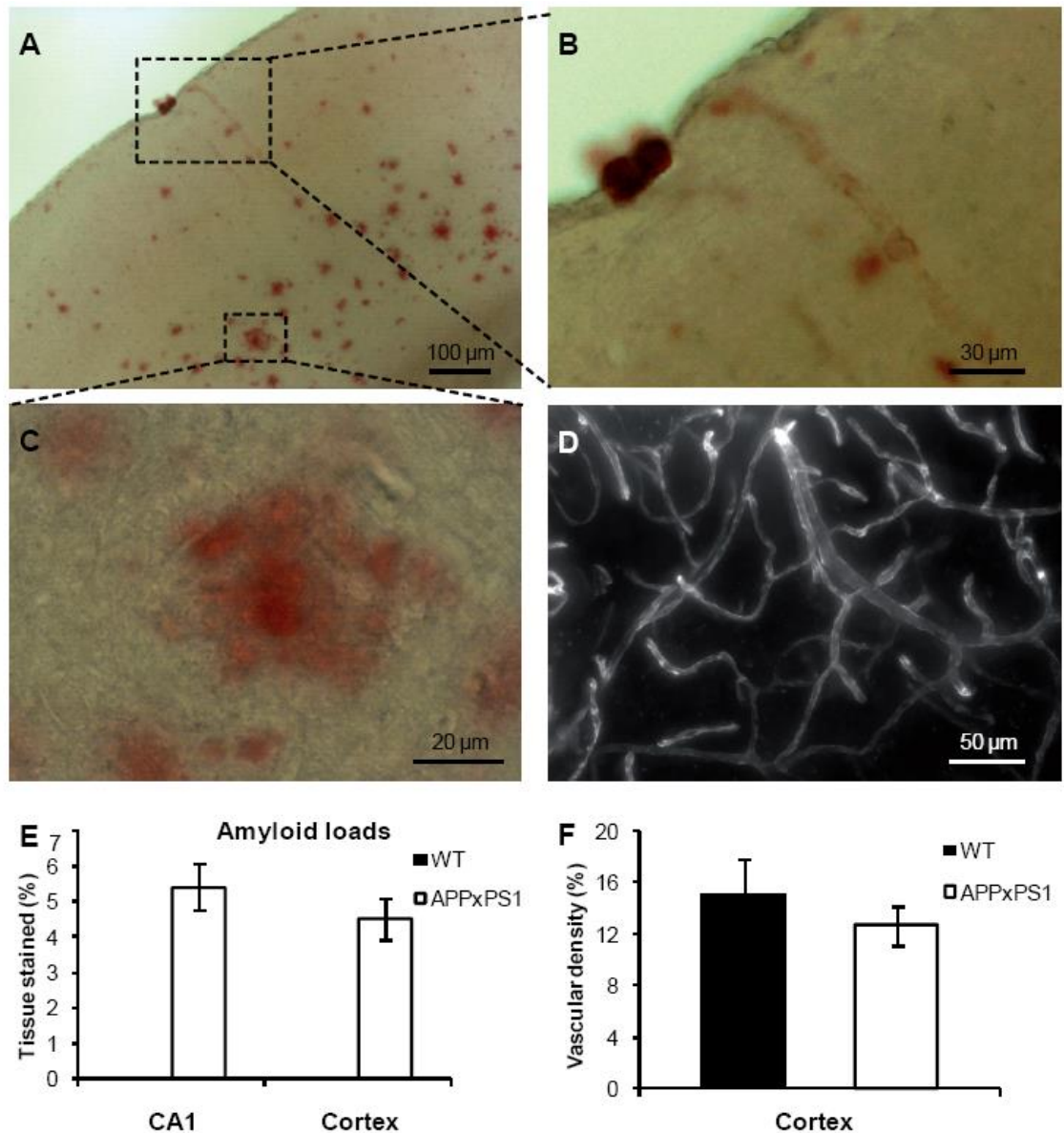

**Figure 1: Neuropathological profile of 8 month-old APPxPS1 mice. A-C,** APPxPS1 develop brain amyloidosis as evidenced by numerous Congo red positive plaques in the parenchyma and blood vessels. B and C are magnifications of a blood vessel positive for Congo red and a parenchymal amyloid deposit, respectively, as outlined in A. **D,** Immunolabbeling of collagen IV revealed blood vessels architecture within the cortex. **E,** Regional amyloid loads (mean ± sem) were calculated from Congo red sections in different brain region (somatosensory cortex and CA1). Note that the relative volume occupied by amyloid loads is not significantly different in CA1 versus Cortex of APPxPS1 mice. **F,** Vascular density (mean ± sem) was calculated from collagen IV immunostaining in the somatosensory cortex.

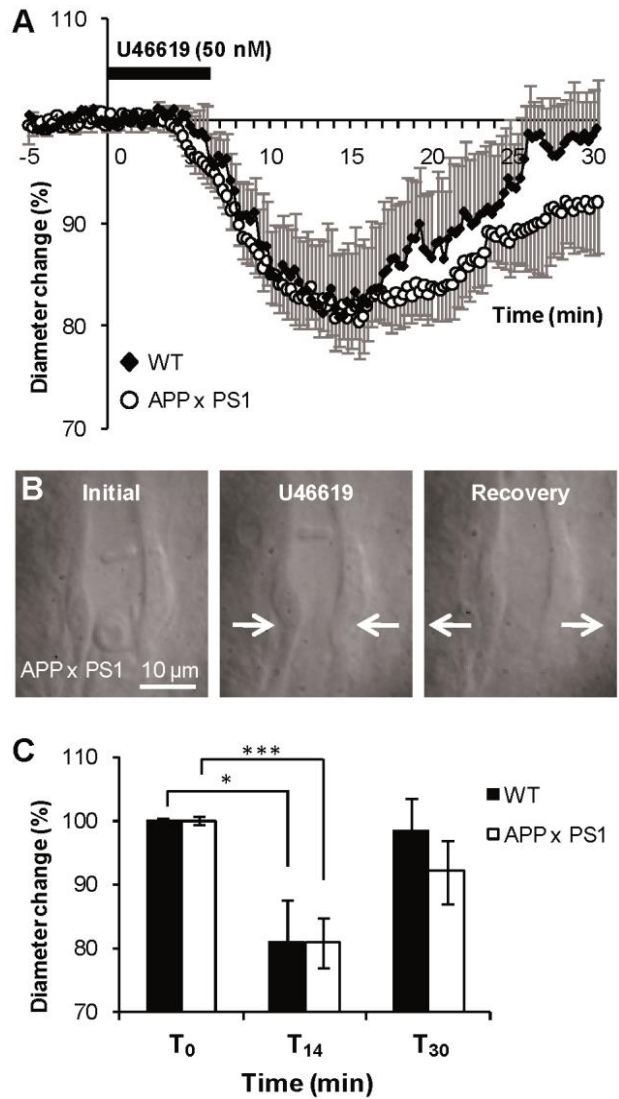

**Figure 2: Vasoconstrictions induced by direct stimulation of blood vessels using U46619 were not significantly different in APPxPS1 compared to WT mice.**

**A**, Mean vascular constrictions ± sem induced by U46619 (50 nM) in 8 month-old APPxPS1 (n = 10, open circles) and WT (n = 8, filled lozenges) mice. **B**, Infrared images of an intracortical blood vessel that reversibly constricted in response to bath application of U46619 (50 nM) in an APPxPS1 mouse. **C**, At 14 min ($T_{14}$), vasoconstrictions shown in (A) were significant in WT ($p<0.05$) as in APPxPS1 ($p<0.001$), compared to values at $T_0$. No significant differences were detected at 14 min, or at 30 min, between these two lines of mice ($p>0.3$).

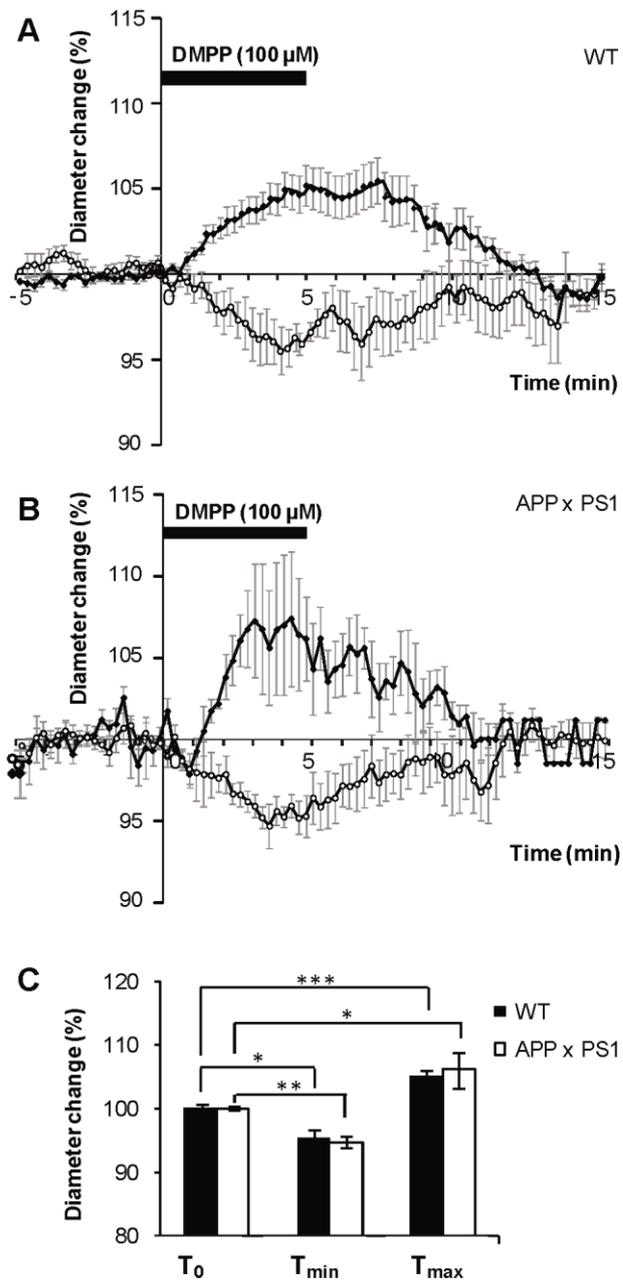

**Figure 3: Vasodilatations and vasoconstrictions induced by indirect blood vessels stimulation, via DMPP induced interneurons activation, were not significantly different in APPxPS1 compared to WT mice. A**, Mean vascular dilatations (n = 12) and constrictions (n = 7) ± sem induced by DMPP (100 µM) in 8 month-old WT mice. **B**, Mean vascular dilatations (n = 3) and constrictions (n = 4) ± sem induced by DMPP (100 µM) in 8 month-old transgenic APPxPS1 mice. **C**, Mean maximal or minimal responses for vasodilatation and vasoconstriction, respectively, were compared to initial values (at $T_0$) for WT and APPxPS1 mice and between WT and APPxPS1 mice.

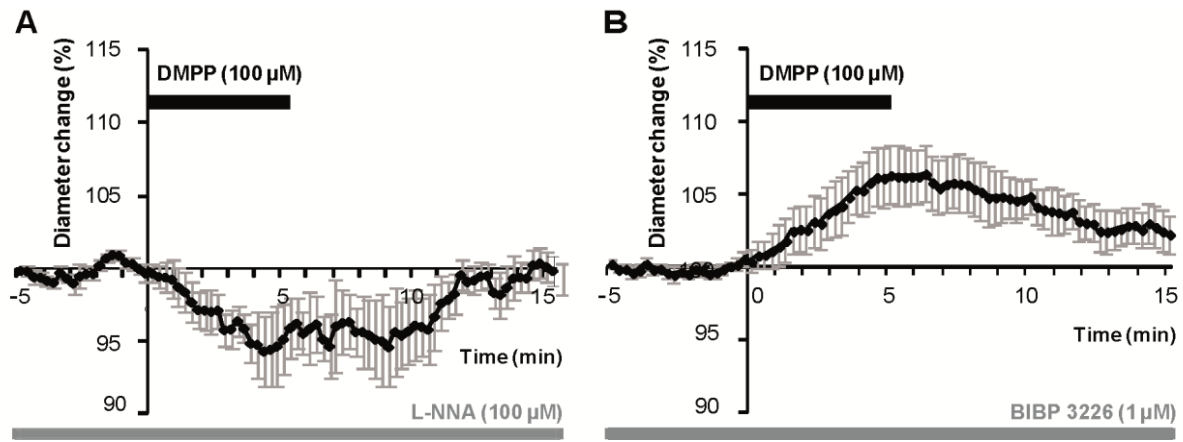

**Figure 4: DMPP induced vasodilations are mediated by NO while constrictions are mediated by NPY in 8 month-old WT mice. A,** Mean vasoconstriction (n = 7) ± sem induced by DMPP (100 µM) in the presence of nNOS inhibitor L-NNA (100 µM). **B,** Mean vascular dilation (n = 5) ± sem induced by DMPP (100 µM) in the presence of the NPY Y1 receptor antagonist BIBP 3226 (1 µM).

**Table 1**

Vascular reactivity induced by U46619 (50 nM) or DMPP (100 µM) applications in WT and APPxPS1 mice cortical slices. ~~Mean v~~Vascular dilation (white line) or constriction (grey lines) are indicated for each condition as mean diameter change or mean AUC ± sem. Proportions of reactive of blood vessels following ~~by~~ U46619 (50 nM) or DMPP (100 µM) applications were determined in WT and APPxPS1 mice cortical slices.

|   | Diameter change (%) | | AUC | | Reactivity (%) | |
|---|---|---|---|---|---|---|
|   | **WT** | **APPxPS1** | **WT** | **APPxPS1** | **WT** | **APPxPS1** |
| **U46619** | 80.18 ± 4.79 | 80.83 ± 3.79 | 183.94 ± 37.91 | 325.15 ± 47.31 | 57 | 56 |
| **DMPP** | 105.21 ± 1.21 | 106.08 ± 2.91 | 39.40 ± 8.49 | 39.22 ± 21.23 | 44 | 15 |
|   | 95.52 ± 1.08 | 94.71 ± 0.88 | 27.77 ± 9.64 | 35.86 ± 14.89 | 26 | 20 |

**Table 2**

Vascular reactivity induced by DMPP applications in WT mice cortical slices in the presence of L-NNA or BIBP 3226. Vascular dilation (white line) or constriction (grey lines) are indicated for each condition as mean diameter change or mean AUC ± sem. Proportions of reactive of blood vessels following DMPP (100 µM) applications were determined in WT mice.

|   | Diameter change (%) | AUC | Reactivity (%) |
|---|---|---|---|
|   | **WT** | **WT** | **WT** |
| **DMPP + BIBP 3226** | 106.24 ± 2.02 | 39.75 ± 6.21 | 38 |
| **DMPP + L-NNA** | 94.65 ± 2.41 | 26.76 ± 6.00 | 44 |